\documentclass[preprint]{sigchi}

\doi{}
\isbn{}

\usepackage{balance}  
\usepackage{graphics} 
\usepackage{enumitem}
\usepackage[T1]{fontenc}
\usepackage{txfonts}
\usepackage{mathptmx}
\usepackage[pdftex]{hyperref}
\usepackage{color}
\usepackage{multirow}
\usepackage{booktabs}
\usepackage{textcomp}
\usepackage{microtype} 
\usepackage{ccicons}  

\usepackage{todonotes}

\usepackage{xspace}

\def\plaintitle{Beagle: Automated Extraction and Interpretation of Visualizations from the Web}

\def\emptyauthor{}
\def\plainkeywords{Authors' choice; of terms; separated; by
  semicolons; include commas, within terms only; required.}

\makeatletter
\def\url@leostyle{%
  \@ifundefined{selectfont}{
    \def\UrlFont{\sf}
  }{
    \def\UrlFont{\small\bf\ttfamily}
  }}
\makeatother
\def\pprw{8.5in}
\def\pprh{11in}

\definecolor{linkColor}{RGB}{6,125,233}
\hypersetup{%
  pdftitle={\plaintitle},
  pdfauthor={\emptyauthor},
  pdfkeywords={\plainkeywords},
  bookmarksnumbered,
  pdfstartview={FitH},
  colorlinks,
  citecolor=black,
  filecolor=black,
  linkcolor=black,
  urlcolor=linkColor,
  breaklinks=true,
}

\newcommand{\sys}{Beagle\xspace}
\newcommand{\ie}{\emph{i.e.,}\xspace}
\newcommand{\eg}{\emph{e.g.,}\xspace}
\newcommand{\etal}{\emph{et~al.}\xspace}

\newif\ifnotes
\notestrue

\usepackage[normalem]{ulem}
\definecolor{mred}{rgb}{.80,.12,.30}
\definecolor{grey}{rgb}{0.5,0.5,0.5}

\let\origcite\cite
\renewcommand{\cite}[1]{\ifnotes\mbox{\origcite{#1}}\else \origcite{#1}\fi}

\begin{document}

\title{\plaintitle}

\numberofauthors{3}
\author{%
  \alignauthor{Leilani Battle\\
    \affaddr{University of Washington}\\
    \affaddr{Seattle, USA}\\
    \email{leibatt@cs.washington.edu}}\\
  \alignauthor{Peitong Duan\\
    \affaddr{Google}\\
    \affaddr{Mountain View, USA}\\
    \email{peitongd@google.com}}\\
  \alignauthor{Zachery Miranda\\
    \affaddr{MIT}\\
    \affaddr{Cambridge, USA}\\
    \email{zmiranda@mit.edu}}\\
  \alignauthor{Dana Mukusheva\\
    \affaddr{MIT}\\
    \affaddr{Cambridge, USA}\\
    \email{d.mukusheva@gmail.com}}\\
  \alignauthor{Remco Chang\\
    \affaddr{Tufts University}\\
    \affaddr{Medford, USA}\\
    \email{remco@cs.tufts.edu}}\\
  \alignauthor{Michael Stonebraker\\
    \affaddr{MIT}\\
    \affaddr{Cambridge, USA}\\
    \email{stonebraker@csail.mit.edu}}\\
}

\maketitle

\begin{abstract}
``How common is interactive visualization on the web?'' ``What is the most popular visualization design?'' ``How prevalent are pie charts \emph{really}?'' These questions intimate the role of interactive visualization in the real (online) world. In this paper, we present our approach (and findings) to answering these questions. First, we introduce Beagle, which mines the web for SVG-based visualizations and automatically classifies them by type (i.e., bar, pie, etc.). With Beagle, we extract over 41,000 visualizations across five different tools and repositories, and classify them with 86\% accuracy, across 24 visualization types. Given this visualization collection, we study usage across tools. We find that most visualizations fall under four types: bar charts, line charts, scatter charts, and geographic maps. Though controversial, pie charts are relatively rare in practice. Our findings also indicate that users may prefer tools that emphasize a succinct set of visualization types, and provide diverse expert visualization examples.

\end{abstract}



\section{Introduction}

The Visualization community has made tremendous strides in the past decade or so in bringing data visualization to the masses, and recently in creating and sharing visualizations online. Tools such as Tableau, D3, Plotly, Exhibit and Fusion Charts have made the design and publication of interactive visualizations on the web easier than ever. Popular venues such as the New York Times, the Guardian, and Scientific American then utilize these tools as an effort to democratize data, resulting in unprecedented advances in data journalism, visual storytelling, and browser-based rendering techniques.

However, while it is generally believed that visualization has reached the public, it remains unclear just how wide the reach is. Although the Visualization research community continues to design and develop new interactive visualization techniques (particularly for the web), we still have little idea as to how frequently these techniques are used.
For example, many have advocated for the ``death'' of pie charts~\cite{nussbaumer_death_2011,few2007save,Tufte1986VDQ}, but just how commonly used are pie charts compared to the recommended alternative, the bar chart?
 We aim to investigate these important questions in the context of rendering and publishing visualizations on the web.

In this paper, we present our approach and findings to these questions for web-based visualizations. In order to mine and classify visualizations on the web, we developed \sys{}. \sys is an automated system to extract SVG-based visualizations rendered in the browser, label them, and make them available as a query-able data store. \sys consists of two major components: a Web Crawler for identifying and extracting SVG-based visualizations from web pages, and an Annotator for automatically classifying extracted visualizations with their corresponding visualization type.

To date, we have used \sys{} to extract visualizations from five different web-based visualization tools and repositories, totaling over 41,000 visualizations. 
We evaluate \sys{}'s classification accuracy using our extracted visualization collections. We find that \sys{} can correctly classify SVG-based visualizations with 86\% accuracy, in a multi-class classification test across 24 visualization types observed on the web.

In an analysis of our visualization collections, we find that the vast majority of visualizations fall under \emph{only four types}: bar charts, line charts, scatter charts, and geographic maps. We also find that the most popular visualizations varied across visualization tools, indicating that some tools may be more accessible or appealing for certain visualization types. Despite this variation, bar and line charts clearly dominate usage of the visualization tools and repositories that we studied.

Except for D3, we only observed 14 or fewer visualization types for each tool, pointing to a possible barrier in knowledge transfer from research to general use.

These results may indicate that even though they lack visualization types from a research perspective, current tools may have an excess of visualization types for real-world users. When extracting the D3 collection, we found that users often copied established D3 examples as a starting point for creating new visualizations, showing promise as a technique to encourage users to try out new visualization types.

Unlike other mediums, such as presentations and reports, we find that pie charts are relatively rare in the extracted collections. As such, the controversy surrounding pie charts appears to be a moot point for web-driven visualizations.

To summarize, we make the following contributions:
\begin{itemize}[topsep=0pt, partopsep=0pt, itemsep=-2pt, leftmargin=*]
\item{a new technique for classifying SVG objects. When performing five-fold cross validation, our classification technique provides: 1) 82\%-99\% classification accuracy \emph{within} collections, and 2) 86\% accuracy \emph{between} collections.}
\item{we analyzed the result of mining over 41,000 pages that contain SVG visualizations, out of roughly 20 million pages that were visited. We found that interactive visualizations still represent a small number of web pages on the internet. We present our analysis results in the Section titled ``Discussion: Visualization Usage on the Web''.}
\end{itemize}

\vspace{-2mm}
\section{Finding and Extracting Visualizations}
\label{sec:datacollection}

In this section, we describe how we used our Web Crawler to collect thousands of visualizations from the web.

We initially performed a general, unguided crawl from the web to discover visualizations. However, after crawling 20 million webpages, we only found roughly 10,000 pages with visualizations, or 0.05\%. The majority of these webpages were user profile pages from stackoverflow.com (and stackexchange websites), each with a single line chart showing user activity over time, resulting in thousands of redundant visualizations.

As a result, we sought out specific ``islands'' of usage on the web, where users frequently deposit their visualizations at a single, centralized source website. We performed a broad search for islands that consistently contained SVG-based visualizations, investigating sites such as Tableau Public~\cite{tableau_software_tableau_2016}, Many Eyes~\cite{viegas_manyeyes:_2007}, and bl.ocks.org~\cite{bostock_popular_2016} for D3 examples~\cite{bostock_d3_2011}. After filtering out the islands that use raster formats instead of SVG, five islands were successfully mined using our Web Crawler: bl.ocks.org, Plotly~\cite{plotly_plotly_2016}, Chartblocks~\cite{noauthor_chartblocks}, Fusion Charts~\cite{noauthor_fusion}, and the Graphiq knowledge base~\cite{noauthor_graphiq}.

For each URL from the visualization islands, the Web Crawler identifies any SVG objects on the corresponding webpage, and extracts the raw SVG specification for each object, as well as a snapshot of the object.

\textbf{Web Crawler Results}:
Using the urls collected from our targeted web search, we ran the Web Crawler to visit the corresponding webpages to extract SVG objects.
We label the resulting collections by their corresponding visualization tool or repository name (D3, Plotly, Chartblocks, Fusion Charts, and Graphiq). Given that our islands are websites dedicated to a specific tool or repository, each url collected from our islands was likely to produce SVG-based visualizations, allowing us to quickly collect thousands of visualizations per run.
The crawls resulted in over 42,000 total SVG-based visualizations. Per island, we found: over 2000 visualizations for D3, over 15000 visualizations for Plotly, over 22000 visualizations for Chartblocks, over 500 visualizations for Fusion Charts, and over 2500 visualizations for Graphiq.

\vspace{-2mm}
\section{Automatically Labeling Visualizations}
\label{sec:annotator}

Labeling visualizations by type helps us gain a sense for how different visualization types are designed and shared on the web. Given that the Web Crawler can extract thousands of visualizations, an automated process is needed to efficiently label visualizations. In this section, we explain how the \sys Annotator uses a new SVG-based classifier to automatically classify extracted visualizations by type.

The Annotator calculates basic statistics over SVG elements to discern visualization types. Examples of some statistics are: the average x position of SVG elements (\eg \texttt{rect}'s, \texttt{circle}s, etc.), average width of SVG elements, and count of unique colors observed. Each basic statistic represents a single classification feature.
The features are then passed to an \emph{off-the-shelf decision tree classifier} (Python's Scikit-Learn with default parameters) to determine the visualization types.

Our feature extraction code computes statistics over the positions, sizes, and basic styles for SVG elements (114 features total). These features cannot be individually covered here due to space constraints. Instead, we summarize the groups of features we extract, and provide intuition for why they were selected. Features belong to one of three groups: general (6 features), stye (19 features), and per-element (89 features). General features measure the occurrence of element \emph{types} in a visualization. Style features track how fill, border, and font styles are applied for all elements in the visualization, regardless of element type. Per-element features track the positions, sizes, and styles of five element types: \texttt{circle} (16), \texttt{rect} (20), \texttt{line} (15), \texttt{path} (35), and \texttt{text} (3). For the rest of this section, we explain how we calculate each feature group.

\vspace{-2mm}
\subsubsection{General features}
The intuition behind the general features is to summarize the prevalence of certain elements within each visualization. This may be an early indicator for certain visualization types. For example, heavy use of circles may indicate a scatter or bubble chart. These features consist mainly of counts for the five SVG element types. For each SVG object (\ie each extracted visualization), we count the instances of each element type.

We also count the number of horizontal and vertical axis lines, to differentiate visualizations that often contain axes (\eg bar charts), from ones that do not (\eg geographic maps).

\vspace{-2mm}
\subsubsection{Style features}
Differences in visualization types can be found in the way that SVG elements are styled. For example, how they are colored, and given border and line thicknesses. Styling is often treated as orthogonal to the layout of elements in the DOM tree, so we analyze styling using separate features.

Counting the unique colors observed can be useful, because many visualizations have color applied in a systematic way (\eg all bars in a bar chart are one color). We count the number of unique fill and border colors as two separate features. We also consider the maximum and minimum stroke widths, as well as maximum and minimum font size. Given that text can vary widely in font size, such as for word clouds, we calculate the number of unique font sizes, and font size variance.

Note that we also take into account the inheritance of styles through parent objects and embedded CSS style properties.

\vspace{-2mm}
\subsubsection{Per-Element features}
We analyze each SVG element type to further differentiate visualization designs. For example, rect elements with the same y position might indicate a bar chart, whereas circles with identical radii might indicate a scatterplot. Note that all features are normalized: x positions are divided by visualization width, y positions by visualization height, all line and path lengths by visualization diagonal, and all shape widths by either width or height (whichever is larger).

\textit{Features extracted for all elements}:
For all element types, we calculate a standard set of statistics. First, we calculate the maximum, minimum, variance, and total unique x positions, and then repeat for y position. We then calculate the average number of elements that share positions, and the total unique CSS class names.
Tracking x and y positions allows us to identify layout patterns. For example, vertical bar charts have rectangles with identical y position, and periodic x positions. In contrast, scatter charts have circles with many unique x and y positions, helping to discern bar charts from scatter charts.

\textit{Circle Features}:
We calculate the maximum, minimum, and variance in the radii. We also consider the maximum number of circles with identical radii. Radii variance helps to discern visualizations with equal-size circles (\eg scatter charts), from those with varying circle sizes (\eg bubble and radial charts).

\textit{Rect Features}:
We calculate: the maximum, minimum, and variance in rect widths; the maximum number of rects with identical width; and the number of unique widths observed. We repeat these calculations for rect heights. Similar to positioning, tracking widths and heights helps to identify size patterns, such as equal-width or equal-height bars in bar charts.

\textit{Path Features}:
We mainly consider the total characters used to specify a path. To do this, we analyze the ``d'' attribute, which contains commands for drawing the path (\eg move to point A, draw a line to point B, etc.). We calculate the maximum, minimum, mean and variance in the length of the ``d'' attribute. This feature is useful for paths, because the longer a ``d'' attribute is, the more detailed it is. Complex shapes require detailed paths, such as countries in geographic maps. We also calculate the Euclidean distance between a path's start and end points, to help discern short paths (\eg parallel coordinates) from long paths (\eg line charts). 

However, path elements are complex, and can be used in place of other SVG elements, requiring some additional statistics. To find polygon-heavy visualizations (\eg voronoi visualizations), we compute the above statistics specifically for paths that contain polygons (\ie paths that start and end in the same place).
We also compute the number of arc calls within a path element, which can help to identify circles in path elements.

\textit{Line Features}:
For lines, we calculate the maximum, minimum, and variance in line length.

\section{Accuracy Results for Labeling Visualizations}
One of \sys{}'s important features is its ability to automatically label visualizations. As such, it is necessary to measure \sys{}'s accuracy, to ensure that people can rely on the classification labels. In this section, we evaluate \sys{}'s SVG-based classifier in two ways. First we perform a ``within-group'' evaluation where we trained and evaluated the classifier using the five visualization collections extracted by the \sys Web Crawler. Second, we conduct a ``between-group'' evaluation where we randomly mixed visualizations from the different collections and evaluated the accuracy of the classifier.

\vspace{-2mm}
\subsection{Within-Group Evaluation}

\begin{table*}[]
\centering
\begin{tabular}{|l|l|l|l|}
\hline
Collection       & Size & Total Types & Visualization Type Labels                                                                                                                                                                                                                                                                                                                                        \\ \hline
D3      & 1247 & 22          & \begin{tabular}[c]{@{}l@{}}area (32), bar (154), box (11), bubble (70), chord (34), donut (31), heatmap (32),\\ \textbf{geographic map (379)}, graph (60), hexabin (21), line (157), radial (13), pie (7),\\ sankey (11), scatter (118), treemap (10),voronoi (25),waffle (12),word cloud (6),\\ sunburst (28), stream graph (13), parallel coordinates (23)\end{tabular} \\ \hline

Chartblocks     & 22730 & 4          & \begin{tabular}[c]{@{}l@{}}pie (5514), \textbf{line (8065)}, bar (7402), scatter (1749)\end{tabular} \\ \hline

Fusion Charts      & 530 & 10          & \begin{tabular}[c]{@{}l@{}}area (14), \textbf{bar (224)}, box (22), donut (54), geographic map (48), heatmap (12),  line (84), \\ pie (26), scatter (29), sunburst (6)\end{tabular} \\ \hline

Graphiq      & 2727 & 11          & \begin{tabular}[c]{@{}l@{}}bubble (9), donut (18), area (210), graph (5), geographic map (244), line (655), waffle (4), \\ box (6), \textbf{bar (1542)}, treemap (6), scatter (28)\end{tabular} \\ \hline

Plotly           & 6544 & 11          & \begin{tabular}[c]{@{}l@{}}area (10), bar (1364), box (259), contour (118), donut (193), filled-line (126),\\ geographic map (184), line (1198), pie (26), radial (17), \textbf{scatter (3049)}\end{tabular}                                                                                                                                                              \\ \hline
\end{tabular}
\caption{General information about each visualization collection in our evaluation.}
\label{tab:dataset:info}
\end{table*}

We use all five of our visualization collections for our validation experiments. Table~\ref{tab:dataset:info} provides a summary of these details for each visualization collection. Each collection was created by mining the corresponding website using the \sys Web Crawler (see Section~\ref{sec:datacollection} for more details). The collections range in size from 530 visualizations (Fusion Charts), up to 23,270 visualizations (Chartblocks), and from 4 visualization types in one collection (Chartblocks) to 22 types (D3).

Note that for our analyses, we omit roughly one quarter of the visualizations extracted by the Web Crawler. This is because the corresponding webpages contained complex features that interfered with the extraction process, such as animations.

\vspace{-2mm}
\subsubsection{Dataset Labels}
We apply classification labels to every visualization used in our analyses (\ie the visualization type).
We considered the visualization types that appeared in all of our data collections, and created a single set of labels to cover them. The final set of labels is provided in Table~\ref{tab:dataset:info}, along with the number of samples observed for each visualization type and each data collection. Visualizations were labeled by the authors either by hand (for bl.ocks.org/D3, Fusion Charts, and Graphiq) or through code (for Plotly and Chartblocks).

We found strong similarities between certain visualization types within the following two groups, and consolidated them: geographic maps (e.g., choropleth and map projections), and graphs and trees (e.g., dendrograms and trees). We also consolidated visualization types with style variations. For example, stacked bar charts and grouped bar charts were consolidated.

\vspace{-2mm}
\subsubsection{Experimental Setup and Results}
We calculate the overall accuracy as the fraction of correct answers across ten runs of stratified, five-fold cross validation. With five-fold cross-validation, each dataset is shuffled and partitioned into five groups (or folds). For each fold, the classifier is trained on the other four folds (or 80\% of the data), and tested on the fold that is left out (20\% of the data).

The goal of these experiments is to test \sys{}'s ability to accurately label visualizations across a variety of rendering environments. Here, we have five separate visualization collections, represented by the visualization tools used to create the visualizations  (D3, Plotly, Chartblocks, Fusion Charts, and Graphiq). The results of our experiments are provided in Table~\ref{tab:results:all_results}. We see that in all five cases, \sys provides 83.1\% classification accuracy or higher. In four of five cases, \sys provides 96\%-99\% accuracy. Through these results, we confirm that \sys is able to capture the defining characteristics of different visualization types across collections, simply by calculating basic statistics over the SVG objects.

\subsection{Between-Group Evaluation}
\begin{table}
\centering
\begin{tabular}{|l|l|l|}
\hline
Collection        & \begin{tabular}[c]{@{}l@{}}Training/Test Set\\ Sizes (Per Fold)\end{tabular} & Accuracy \\ \hline
D3                & 80\%/20\%                                                                    & 83.1     \\ \hline
Plotly            & 80\%/20\%                                                                    & 97.1     \\ \hline
Chartblocks       & 80\%/20\%                                                                    & 99.5     \\ \hline
Fusion Charts     & 80\%/20\%                                                                    & 96.3     \\ \hline
Graphiq           & 80\%/20\%                                                                    & 99.2     \\ \hline
Mixture           & 80\%/20\%                                                                    & 86.5     \\ \hline
Mixture + Non-Vis & 80\%/20\%                                                                    & 94.0     \\ \hline
\end{tabular}
\caption{Classification accuracy (in percent) for individual collections, and for combining the collections (Mixture, Mixture + Non-Vis).}
\label{tab:results:all_results}
\end{table}

In the previous section, we found that the \sys Annotator achieves high classification accuracy for visualizations rendered using a particular tool (i.e., classification \emph{within groups}).
However, on the web, \sys has to contend with a mixture of SVG objects, where some may have been created by different tools, and others may not even be visualizations (e.g., logos). Thus, it is important to also test \sys{}'s performance when faced with a mix of different SVG outputs. To test this, we formed a new collection by randomly selecting 500 visualizations from each visualization collection. All visualization types for each collection are represented in the samples. We performed 10 runs of five-fold cross validation on the mixture of 2500 visualizations, and found that \sys provides 86.5\% classification accuracy (labeled as ``Mixture'' in Table~\ref{tab:results:all_results}).

To test how \sys discerns between visualizations and other non-visualization SVG objects, we added 1,000 non-visualizations to the mixture, resulting in 3,500 total SVG objects. We trained a binary classifier using \sys{}'s features, and found that \sys provides 94.0\% classification accuracy (labeled as ``Mixture + Non-Vis'' in Table~\ref{tab:results:all_results}).

We found that \sys is comparable to the reported performance of related classification techniques, such as those used in Revision~\cite{savva_revision:_2011} (80-90\% accuracy), FigureSeer~\cite{siegel_figureseer:_2016} (86\% accuracy), and ChartSense~\cite{Jung2017chartsense} (76.7-91.3\% accuracy). However, an advantage of \sys is its ability to classify different visualization types with a limited number of training examples. Computer vision and deep learning techniques generally require large training sets to develop their classification models, which may not be possible when building a new visualization collection from scratch. Nevertheless, to provide a direct comparison, we also ran the Revision\footnote{Note that Revision requires bitmap images as input.} classifier with our mixed visualization collection (80.7\% accuracy), and found that \sys provides a 5.8\% improvement in classification accuracy.

\begin{table*}[]
\centering
\begin{tabular}{|l|l|l|l|l|l|l|}
\hline
Collection    & All Types Observed & Most Popular    & 2nd Most Popular & \% Bar & \% Line & \% Pie \\ \hline
Chartblocks   & 4                  & line, 34.7\%    & Bar, 31.8\%      & 31.8\% & 34.7\%  & 23.2\% \\ \hline
D3            & 25                 & Map, 30.4\%     & Line, 12.6\%     & 12.3\% & 12.6\%  & 0.6\%  \\ \hline
Fusion Charts & 14                 & Bar, 42.3\%     & Line, 15.8\%     & 42.3\% & 15.8\%  & 4.9\%  \\ \hline
Graphiq       & 12                 & Bar, 56.5\%     & Line, 24.0\%     & 56.5\% & 24.0\%  & 0\%    \\ \hline
Plotly        & 11                 & Scatter, 46.6\% & Bar, 20.8\%      & 20.8\% & 18.3\%  & 0.4\%  \\ \hline
\end{tabular}
\caption{Relevant totals for visualization types across the visualization collections (raw data for columns 3-7 available in Table~\ref{tab:dataset:info}).}
\label{tab:discussion:numbers}
\end{table*}

\section{Discussion: Visualization Usage on the Web}
\label{sec:discussion}

\subsection{Supporting Fewer Visualization Types is Common}
Here, we discuss our observations in the diversity versus usage of visualization types. For some collections, we observed more visualization types than are represented in our evaluation (i.e., Table~\ref{tab:dataset:info}). For example, we observed exactly one waffle chart from our crawl of the Fusion Charts website. The total observed visualization types are provided in Table~\ref{tab:dataset:info}. Except for D3, the other visualization collections have limited variety in visualization types. This could be due to a disconnect between visualization research, where there are significantly more than 14 visualization types, and general practice (i.e., what we have observed across multiple visualization tools). Furthermore, we find that the more obscure visualization types generally: 1) illustrate complex relationships between data groups (\eg graphs, sunbursts, treemaps, and arc diagrams); or 2) have more popular (and often simpler) equivalents, such as waffle charts and word clouds, which could be replaced by bar charts. These findings suggest that when more complex visualization types are supported, users may generally avoid them, which may also contribute to the low number of supported visualization types.

\subsection{Line and Bar Charts Dominate the Collections}
We see in Table~\ref{tab:discussion:numbers} that the most popular visualization type varies across visualization collections, with a steep drop in usage between the most popular and second most popular visualization types.
We found geographic maps, line charts, bar charts, and scatter charts to be the most popular types across the visualization collections.
Together, these four visualization types represent a large fraction of the visualization collections: 64.8\% for D3, 72.6\% for Fusion Charts, 88.6\% for Plotly, and 90.5\% for Graphiq.
Furthermore, line and bar charts are always within the top 3 most popular visualization types across all collections (usage reported in columns 5 and 6 of Table~\ref{tab:discussion:numbers}).

As such, it seems that just adding more visualization types to the tools may result in only small changes to how users choose to display their data. However, having easy-access examples that are well documented, such as the D3.js visualization gallery and tutorials pages, could have an effect. Throughout the course of our crawl of the bl.ocks.org website (and subsequent analysis), we did find that many visualizations created by users were very similar to the available D3 examples.

\subsection{Pie Charts are Not Popular}
Interestingly, though pie charts are generally considered to be a well-known and relatively simple visualization type, they represent only a small fraction of the visualizations that we observed on the web. For four of the five visualization tools and repositories that we studied, pie charts represent less than 5\% of the visualizations we observed. The only exception was for Chartblocks, where only four visualization types are observed (line, bar, scatter, and pie). In this case, pie charts are 23.2\% of the collection. However, even in the case of Chartblocks, we find that line charts are 34.7\% of the collection (12\% higher than pie charts), and bar charts are 31.8\% of the collection (8.6\% higher than pie charts), again demonstrating (in this case a slight) preference for these visualization types over pie charts. Thus when given multiple options for visualization type, people seem to avoid using pie charts. It is unclear whether this avoidance stems from existing visualization conventions (e.g., \cite{nussbaumer_death_2011,few2007save,Tufte1986VDQ}), or other factors, such as the design or presentation of the visualization tools.

\vspace{-2.2mm}
\section{Related Work}

Our goal is to better understand how different tools are utilized to create and share visualizations on the web. In pursuit of this goal, we developed visualization extraction and classification techniques to harvest a collection of visualizations for analysis. In this section, we highlight related projects in these areas.

\sys was inspired by existing projects to mine and analyze website designs, such as Webzeitgeist \cite{kumar_webzeitgeist:_2013} and D.Tour \cite{ritchie_d.tour:_2011}. Some projects within the visualization community also incorporate web mining and visualization extraction, but for different goals. For example, Harper and Agrawala extract data from D3 visualizations to re-style visualizations~\cite{harper_deconstructing_2014}, as well as to generate visualization templates~\cite{harper2017templates}. Saleh~\etal extract infographics from Flickr to develop and evaluate techniques for similarity-based search of infographics~\cite{saleh_learning_2015}.

\sys also extends existing work in visualization classification~\cite{shao_recognition_2006,savva_revision:_2011, prasad_classifying_2007, huang_system_2007,siegel_figureseer:_2016,lee2016viziometrics,Jung2017chartsense}. Most projects focus on classifying raster images of charts \cite{savva_revision:_2011, prasad_classifying_2007, huang_system_2007,siegel_figureseer:_2016,lee2016viziometrics,Jung2017chartsense}. Prasad~\etal \cite{prasad_classifying_2007} and Savva~\etal \cite{savva_revision:_2011} apply computer vision techniques to extract features from raster images for classification. Huang and Tan \cite{huang_system_2007} extract graphical marks from raster images, and use them to classify visualizations. Jung~\etal~\cite{Jung2017chartsense} and Siegel~\etal~\cite{siegel_figureseer:_2016} apply deep learning techniques to classify raster visualizations. Poco and Heer revisit the approach of raster image classification from a new perspective by extracting visualization \emph{specifications} from images~\cite{poco2017reverse}, instead of visualization types. Shao and Futrelle \cite{shao_recognition_2006} analyze vector images from PDF's to classify images across five visualization types. \sys{} calculates basic statistics over SVG as classification features. \sys{}'s classification features apply to any SVG image, and can accurately classify thousands of visualizations, and as many as 24 different visualization types.

Surprisingly, though many techniques have been proposed to classify visualization images, few projects move beyond the extraction and classification phase to perform broader studies of visualization usage, particularly on the web. Here, we highlight two relevant projects. Benson and Karger study usage of the Exhibit tool for publishing data online~\cite{benson_end-users_2014}, and also find that authors' design decisions are influenced by available design examples. In a related area, Lee~\etal developed a platform for extracting and analyzing visualizations from scientific papers~\cite{lee2016viziometrics}, and find that papers with higher densities of images tend to have higher research impact (visualizations comprise 35\% of these images). In contrast, we find that only a small fraction of webpages contain SVG-based visualizations.

\section{Conclusion}
In this paper, we presented \sys{}, an automated system for collecting, labeling and analyzing visualizations created on the web. \sys supports a flexible design with two stand-alone components. The Web Crawler extracts SVG-based visualizations from webpages, and was able to extract over 41,000  visualizations from the web. The Annotator uses SVG-focused classification techniques to label visualizations, and achieves 86\% classification accuracy, in a multi-class classification test with 24 different visualization types.
We then use the resulting visualization collection to study usage of the different visualization types across tools.
We found that only a small fraction of webpages (0.05\%) contain visualizations created using browser-based tools.
Furthermore, the vast majority of visualizations in the collection are covered by just four visualization types: bar charts, line charts, scatter charts, and geographic maps. And though they are hotly debated in the visualization community, pie charts are somewhat rare in practice. Our findings indicate that real-world users may be better served by visualization tools that focus on supporting a small set of visualization types, with clear examples from tool developers for how different visualization types are used.



\bibliographystyle{SIGCHI-Reference-Format}
\bibliography{references}


\begin{thebibliography}{00}


\ifx \showCODEN    \undefined \def \showCODEN     #1{\unskip}     \fi
\ifx \showDOI      \undefined \def \showDOI       #1{{\tt DOI:}\penalty0{#1}\ }
  \fi
\ifx \showISBNx    \undefined \def \showISBNx     #1{\unskip}     \fi
\ifx \showISBNxiii \undefined \def \showISBNxiii  #1{\unskip}     \fi
\ifx \showISSN     \undefined \def \showISSN      #1{\unskip}     \fi
\ifx \showLCCN     \undefined \def \showLCCN      #1{\unskip}     \fi
\ifx \shownote     \undefined \def \shownote      #1{#1}          \fi
\ifx \showarticletitle \undefined \def \showarticletitle #1{#1}   \fi
\ifx \showURL      \undefined \def \showURL       #1{#1}          \fi

\bibitem{benson_end-users_2014}
{Edward Benson} {and} {David~R. Karger}. 2014.
\newblock \showarticletitle{End-users {Publishing} {Structured} {Information}
  on the {Web}: {An} {Observational} {Study} of {What}, {Why}, and {How}}. In
  {\em Proceedings of the {SIGCHI} {Conference} on {Human} {Factors} in
  {Computing} {Systems}} {\em ({CHI} '14)}. ACM, New York, NY, USA, 1265--1274.
\newblock
\showISBNx{978-1-4503-2473-1}
\showDOI{%
\url{http://dx.doi.org/10.1145/2556288.2557036}}


\bibitem{bostock_popular_2016}
{Michael Bostock}. 2016.
\newblock Popular {Blocks} - bl.ocks.org.
\newblock   (Sept. 2016).
\newblock
\showURL{%
\url{http://bl.ocks.org/}}


\bibitem{bostock_d3_2011}
{Michael Bostock}, {Vadim Ogievetsky}, {and} {Jeffrey Heer}. 2011.
\newblock \showarticletitle{D3 {Data}-{Driven} {Documents}}.
\newblock {\em IEEE Transactions on Visualization and Computer Graphics\/}
  {17}, 12 (Dec. 2011), 2301--2309.
\newblock
\showISSN{1077-2626}
\showDOI{%
\url{http://dx.doi.org/10.1109/TVCG.2011.185}}


\bibitem{noauthor_chartblocks}
{Chartblocks}. 2017.
\newblock Online {Chart} {Builder} - {ChartBlocks}.
\newblock   (2017).
\newblock
\showURL{%
\url{https://www.chartblocks.com/en/}}


\bibitem{few2007save}
{Stephen Few} {and} {Perceptual Edge}. 2007.
\newblock \showarticletitle{Save the pies for dessert}.
\newblock {\em Visual Business Intelligence Newsletter\/} (2007), 1--14.
\newblock


\bibitem{harper_deconstructing_2014}
{Jonathan Harper} {and} {Maneesh Agrawala}. 2014.
\newblock \showarticletitle{Deconstructing and {Restyling} {D}3
  {Visualizations}}. In {\em Proceedings of the 27th {Annual} {ACM} {Symposium}
  on {User} {Interface} {Software} and {Technology}} {\em ({UIST} '14)}. ACM,
  New York, NY, USA, 253--262.
\newblock
\showISBNx{978-1-4503-3069-5}
\showDOI{%
\url{http://dx.doi.org/10.1145/2642918.2647411}}


\bibitem{harper2017templates}
{J. Harper} {and} {M. Agrawala}. 2017.
\newblock \showarticletitle{Converting Basic D3 Charts into Reusable Style
  Templates}.
\newblock {\em IEEE Transactions on Visualization and Computer Graphics\/}
  {PP}, 99 (2017), 1--1.
\newblock
\showISSN{1077-2626}
\showDOI{%
\url{http://dx.doi.org/10.1109/TVCG.2017.2659744}}


\bibitem{huang_system_2007}
{Weihua Huang} {and} {Chew~Lim Tan}. 2007.
\newblock \showarticletitle{A {System} for {Understanding} {Imaged}
  {Infographics} and {Its} {Applications}}. In {\em Proceedings of the 2007
  {ACM} {Symposium} on {Document} {Engineering}} {\em ({DocEng} '07)}. ACM, New
  York, NY, USA, 9--18.
\newblock
\showISBNx{978-1-59593-776-6}
\showDOI{%
\url{http://dx.doi.org/10.1145/1284420.1284427}}


\bibitem{noauthor_graphiq}
{Graphiq Inc.} 2017.
\newblock Graphiq {\textbar} {Knowledge} {Delivered}.
\newblock   (2017).
\newblock
\showURL{%
\url{https://www.graphiq.com/}}


\bibitem{Jung2017chartsense}
{Daekyoung Jung}, {Wonjae Kim}, {Hyunjoo Song}, {Jeong-in Hwang}, {Bongshin
  Lee}, {Bohyoung Kim}, {and} {Jinwook Seo}. 2017.
\newblock \showarticletitle{ChartSense: Interactive Data Extraction from Chart
  Images}. In {\em Proceedings of the 2017 CHI Conference on Human Factors in
  Computing Systems} {\em (CHI '17)}. ACM, New York, NY, USA, 6706--6717.
\newblock
\showISBNx{978-1-4503-4655-9}
\showDOI{%
\url{http://dx.doi.org/10.1145/3025453.3025957}}


\bibitem{kumar_webzeitgeist:_2013}
{Ranjitha Kumar}, {Arvind Satyanarayan}, {Cesar Torres}, {Maxine Lim}, {Salman
  Ahmad}, {Scott~R. Klemmer}, {and} {Jerry~O. Talton}. 2013.
\newblock \showarticletitle{Webzeitgeist: {Design} {Mining} the {Web}}. In {\em
  Proceedings of the {SIGCHI} {Conference} on {Human} {Factors} in {Computing}
  {Systems}} {\em ({CHI} '13)}. ACM, New York, NY, USA, 3083--3092.
\newblock
\showISBNx{978-1-4503-1899-0}
\showDOI{%
\url{http://dx.doi.org/10.1145/2470654.2466420}}


\bibitem{lee2016viziometrics}
{Poshen Lee}, {Jevin West}, {and} {Bill Howe}. 2016.
\newblock \showarticletitle{Viziometrics: Analyzing Visual Patterns in the
  Scientific Literature}.
\newblock {\em Journal of the Association for Information Science and
  Technology (JASIST) (in prep)\/} (2016).
\newblock


\bibitem{noauthor_fusion}
{InfoSoft Global~Pvt. Ltd.} 2017.
\newblock {JavaScript} {Charts} for {Web}, {Mobile} \& {Apps}.
\newblock   (2017).
\newblock
\showURL{%
\url{http://www.fusioncharts.com/}}


\bibitem{nussbaumer_death_2011}
{Cole Nussbaumer}. 2011.
\newblock death to pie charts.
\newblock   (July 2011).
\newblock
\showURL{%
\url{http://www.storytellingwithdata.com/blog/2011/07/death-to-pie-charts}}


\bibitem{plotly_plotly_2016}
{Plotly}. 2016.
\newblock Plotly {\textbar} {Make} charts and dashboards online.
\newblock   (2016).
\newblock
\showURL{%
\url{https://plot.ly/}}


\bibitem{poco2017reverse}
{Jorge Poco} {and} {Jeffrey Heer}. 2017.
\newblock \showarticletitle{Reverse-Engineering Visualizations: Recovering
  Visual Encodings from Chart Images}.
\newblock {\em Computer Graphics Forum (Proc. EuroVis)\/} (2017).
\newblock
\showURL{%
\url{http://idl.cs.washington.edu/papers/reverse-engineering-vis}}


\bibitem{prasad_classifying_2007}
{V.~S.~N. Prasad}, {B. Siddiquie}, {J. Golbeck}, {and} {L.~S. Davis}. 2007.
\newblock \showarticletitle{Classifying {Computer} {Generated} {Charts}}. In
  {\em 2007 {International} {Workshop} on {Content}-{Based} {Multimedia}
  {Indexing}}. 85--92.
\newblock
\showDOI{%
\url{http://dx.doi.org/10.1109/CBMI.2007.385396}}


\bibitem{ritchie_d.tour:_2011}
{Daniel Ritchie}, {Ankita~Arvind Kejriwal}, {and} {Scott~R. Klemmer}. 2011.
\newblock \showarticletitle{D.{Tour}: {Style}-based {Exploration} of {Design}
  {Example} {Galleries}}. In {\em Proceedings of the 24th {Annual} {ACM}
  {Symposium} on {User} {Interface} {Software} and {Technology}} {\em ({UIST}
  '11)}. ACM, New York, NY, USA, 165--174.
\newblock
\showISBNx{978-1-4503-0716-1}
\showDOI{%
\url{http://dx.doi.org/10.1145/2047196.2047216}}


\bibitem{saleh_learning_2015}
{Babak Saleh}, {Mira Dontcheva}, {Aaron Hertzmann}, {and} {Zhicheng Liu}. 2015.
\newblock \showarticletitle{Learning {Style} {Similarity} for {Searching}
  {Infographics}}. In {\em Proceedings of the 41st {Graphics} {Interface}
  {Conference}} {\em ({GI} '15)}. Canadian Information Processing Society,
  Toronto, Ont., Canada, Canada, 59--64.
\newblock
\showISBNx{978-0-9947868-0-7}
\showURL{%
\url{http://dl.acm.org/citation.cfm?id=2788890.2788902}}


\bibitem{savva_revision:_2011}
{Manolis Savva}, {Nicholas Kong}, {Arti Chhajta}, {Li Fei-Fei}, {Maneesh
  Agrawala}, {and} {Jeffrey Heer}. 2011.
\newblock \showarticletitle{{ReVision}: {Automated} {Classification},
  {Analysis} and {Redesign} of {Chart} {Images}}. In {\em Proceedings of the
  24th {Annual} {ACM} {Symposium} on {User} {Interface} {Software} and
  {Technology}} {\em ({UIST} '11)}. ACM, New York, NY, USA, 393--402.
\newblock
\showISBNx{978-1-4503-0716-1}
\showDOI{%
\url{http://dx.doi.org/10.1145/2047196.2047247}}


\bibitem{shao_recognition_2006}
{Mingyan Shao} {and} {Robert~P. Futrelle}. 2006.
\newblock \showarticletitle{Recognition and {Classification} of {Figures} in
  {PDF} {Documents}}. In {\em Proceedings of the 6th {International}
  {Conference} on {Graphics} {Recognition}: {Ten} {Years} {Review} and {Future}
  {Perspectives}} {\em ({GREC}'05)}. Springer-Verlag, Berlin, Heidelberg,
  231--242.
\newblock
\showISBNx{978-3-540-34711-8}
\showDOI{%
\url{http://dx.doi.org/10.1007/11767978_21}}


\bibitem{siegel_figureseer:_2016}
{Noah Siegel}, {Zachary Horvitz}, {Roie Levin}, {Santosh Divvala}, {and} {Ali
  Farhadi}. 2016.
\newblock \showarticletitle{{FigureSeer}: {Parsing} {Result}-{Figures} in
  {Research} {Papers}}. In {\em Computer {Vision} --€" {ECCV} 2016} {\em
  (Lecture {Notes} in {Computer} {Science})}. Springer, Cham, 664--680.
\newblock
\showISBNx{978-3-319-46477-0 978-3-319-46478-7}
\showDOI{%
\url{http://dx.doi.org/10.1007/978-3-319-46478-7_41}}


\bibitem{tableau_software_tableau_2016}
{Tableau Software}. 2016.
\newblock Tableau {Public}.
\newblock   (2016).
\newblock
\showURL{%
\url{https://public.tableau.com/}}


\bibitem{Tufte1986VDQ}
{Edward~R. Tufte}. 1986.
\newblock {\em The Visual Display of Quantitative Information}.
\newblock Graphics Press, Cheshire, CT, USA.
\newblock
\showISBNx{0-9613921-0-X}


\bibitem{viegas_manyeyes:_2007}
{Fernanda~B. Viegas}, {Martin Wattenberg}, {Frank van Ham}, {Jesse Kriss},
  {and} {Matt McKeon}. 2007.
\newblock \showarticletitle{{ManyEyes}: {A} {Site} for {Visualization} at
  {Internet} {Scale}}.
\newblock {\em IEEE Transactions on Visualization and Computer Graphics\/}
  {13}, 6 (Nov. 2007), 1121--1128.
\newblock
\showISSN{1077-2626}
\showDOI{%
\url{http://dx.doi.org/10.1109/TVCG.2007.70577}}


\end{thebibliography}

\end{document}